# Diffusion Based Cooperative Molecular Communication in Nano-Networks

Neeraj Varshney, *Student Member, IEEE*, Adarsh Patel, *Student Member, IEEE*, and Aditya K. Jagannatham, *Member, IEEE*

***Abstract*—This work presents a novel diffusion based dual-phase molecular communication system where the source leverages multiple cooperating nanomachines to improve the end-to-end reliability of communication. The Neyman-Pearson Likelihood Ratio Tests are derived for each of the cooperative as well as the destination nanomachines in the presence of multi-user interference. Further, to characterize the performance of the aforementioned system, closed form expressions are derived for the probabilities of detection, false alarm at the individual cooperative, destination nanomachines, as well as the overall end-to-end probability of error. Simulation results demonstrate a significant improvement in the end-to-end performance of the proposed cooperative framework in comparison to multiple-input single-output and single-input single-output molecular communication scenarios in the existing literature.**

## I. Introduction

Nanoscale molecular communication has gained significant prominence in recent times due to its ability to provide novel solutions for various problems arising in biomedical, industrial, and surveillance scenarios. Efficient drug delivery, as described in [1], is one such example of its envisaged potential. Several research efforts [2]–[6] have been devoted to developing models as well as analyzing the performance of diffusion based molecular systems. It has been shown in [7] that the distance between two nanomachines significantly affects the performance of communication, since the molecular concentration decays inversely as the cube of this distance. To overcome this impediment, the work in [8] proposed a novel multiple-input multiple-output (MIMO) molecular communication framework, wherein a noticeable performance improvement has been demonstrated through suitable allocation of molecules among the transmitting nodes. However, the work therein focuses exclusively on applications of transmit/ receive diversity and spatial multiplexing in molecular communication, while not exploring cooperative molecular communication that can significantly enhance the communication range with improvement in the reliability through cooperative diversity.

Some works in the existing literature [9], [10] and the references therein have analyzed the performance of relay-assisted molecular communication systems in terms of their error rate, capacity, etc. However, to the best of our knowledge, none of the works in the existing literature consider dual-phase multiple half-duplex nanomachine assisted cooperative molecular communication in the presence of multi-user interference (MUI) and exploit cooperative diversity, which is the key focus of this work.

Neeraj Varshney, Adarsh Patel, and Aditya K. Jagannatham are with the Department of Electrical Engineering, Indian Institute of Technology Kanpur, Kanpur UP 208016, India (e-mail:{neerajv,adarsh,adityaj}@iitk.ac.in).

Towards this objective, the optimal Likelihood Ratio Test (LRT) based decision rule at each cooperating nanomachine is initially derived, followed by a characterization of its probabilities of detection and false alarm. Subsequently, the optimal fusion rule is derived for the destination nanomachine incorporating also the probabilities of false alarm and detection at each of the cooperating nanomachines. Thus, the framework developed is practically applicable, unlike the selective decode and forward protocol considered in works such as [11], [12] that assume retransmission only when the symbol is accurately decoded at the relay. Finally, closed form expressions are derived for the end-to-end probabilities of detection, false alarm, and probability of error for the cooperative scenario under consideration. Simulation results demonstrate the improved performance of the proposed cooperative nanosystem in comparison to the existing multiple-input single-output (MISO) and single-input multiple-output (SIMO) systems [8]. Further, while this work employs multiple molecule types for the source and cooperating nanomachines, it results in a simplistic linear combining based decision rule and also does not require a large capacity backhaul that is necessitated in [8]. Due to space limitations, detailed derivations of some results are given in a supplemental technical report [13].

## II. Diffusion based Cooperative Molecular Communication System Model

Consider a diffusion based molecular communication system with $N$ cooperating nanomachines $R_1, R_2, \cdots, R_N$ as shown schematically in Fig. 1. Further, it can be noted that the diffusion takes place in 3-dimensional and infinite space. The $N$ intermediate nanomachines cooperate with the source nanomachine to relay its information to the destination nanomachine. The transmitter is modeled as a point source while the the destination node is a perfectly absorbing surface. Further, the relay nodes during reception act as perfect absorbing surfaces while during transmission act as point sources. End-to-end communication between the source and destination nanomachines occurs in two phases. In first phase, the source nanomachine emits either $\mathcal{Q}_0$ molecules in the propagation medium for information symbol 1 generated with a prior probability $\beta$ or remains silent for information symbol 0. In the second phase, the intermediate nanomachines $R_i, i = 1, 2, \cdots, N$ that employ decode-and-forwarding protocol, decode the symbol using the concentration received from the source followed by retransmission to the destination nanomachine using different types of molecules[1]. As described in [14], nanomachines such as eukaryotic cells can be genetically modified to emit different types of molecules. The molecules released by the transmitter nanomachine are assumed to diffuse freely through the propagation medium with Brownian motion and are absorbed by the receiver once they reach it. Furthermore, similar to [8], this work also assumes that the inter symbol interference (ISI) and receiver noise are suppressed and negligible in comparison to the MUI. The effect of ISI can be ignored by considering the symbol duration $T_s$ to be significantly larger than $t_p$, where $t_p$ is the time instant when the channel impulse response has a maximum as can be seen in [15, Fig.3]. Further, since such nanonetworks are expected to operate in a distributed and uncoordinated manner, this work focuses on MUI and assumes ISI, noise effects to be negligible in comparison.

[1]The use of different types of molecules at each transmitting nanomachine allows the cooperative nanomachines to operate either in half or full duplex modes [10].

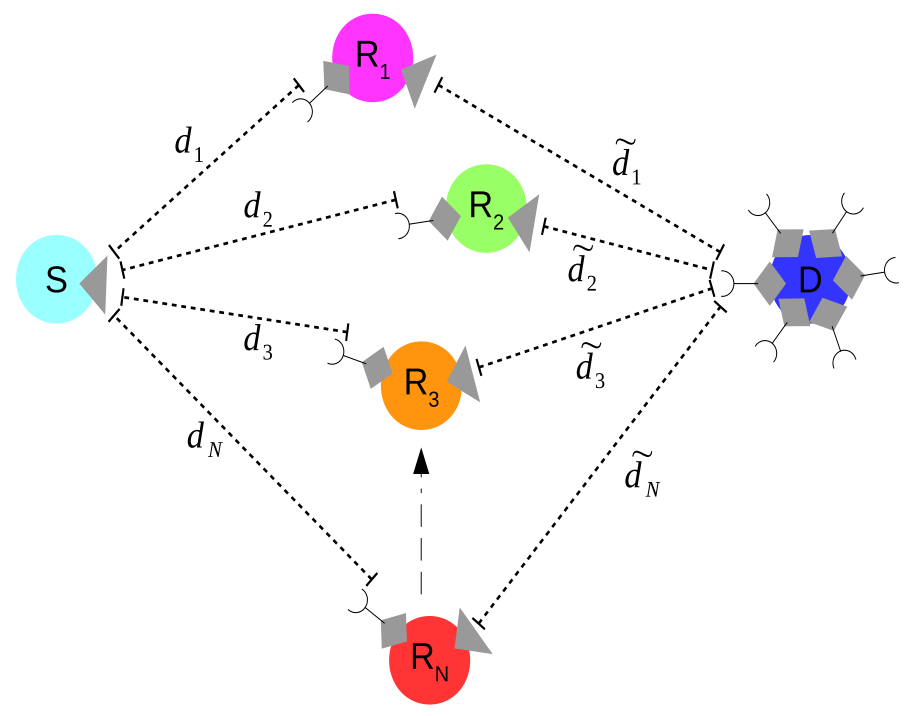

Fig. 1. Schematic diagram of a cooperative molecular nano-network with $d_i$ and $\widetilde{d}_i$ denoting distances from source nanomachine to $R_i$, and $R_i$ to destination nanomachine respectively.

The peak concentration sensed at the $i$th cooperating nanomachine $R_i$ corresponding to transmission by the source nanomachine in the first phase is given as [8]

$$c_i = x_0 \mathcal{Q}_0 h_p(d_i) + \eta_i, \tag{1}$$

where $x_0 \in \{0, 1\}$ denotes the binary information symbol. The term $\eta_i$ represents the MUI at the cooperative node $R_i$. Similar to [8], assuming the MUI to arise from a large number of interfering sources, it can be modeled as a Gaussian random variable with mean $\mu_i$ and variance $\sigma_i^2$, denoted by $\mathcal{N}(\mu_i, \sigma_i^2)$, as per the central limit theorem (CLT) [16]. The molecular concentration is given as

$$\mathcal{Q}_0 h(t) = \mathcal{Q}_0 \frac{1}{(4\pi D t)^{\frac{3}{2}}} \exp\left(-\frac{d_i^2}{4Dt}\right),$$

where $d_i$ denotes the distance between the source and the $i$th node $R_i$. The peak concentration is given at $t = \frac{d_i^2}{6D}$, obtained by solving $\frac{dh(t)}{dt} = 0$. The peak concentration is obtained as

$$\mathcal{Q}_0 h_p(d_i) = \mathcal{Q} d_i^{-3} \left(\frac{3}{2\pi e}\right)^{3/2}. \tag{2}$$

Each cooperative nanomachine employs the minimum probability of error criterion for decoding the transmitted symbol followed by retransmission of the decoded symbol $\widehat{x}_i \in \{0, 1\}$ to the destination nanomachine. Therefore, the peak concentration sensed at the destination corresponding to the $i$th cooperating nanomachine $R_i$ is

$$\widetilde{c}_i = \widehat{x}_i \mathcal{Q}_i h_p(\widetilde{d}_i) + \widetilde{\eta}_i, \tag{3}$$

where $\widetilde{\eta}_i \sim \mathcal{N}(\widetilde{\mu}_i, \widetilde{\sigma}_i^2)$ represents the MUI at the destination nanomachine and $\mathcal{Q}_i$ is the number of molecules emitted by node $R_i$ corresponding to information symbol 1. Similar to $\mathcal{Q}_0 h_p(d_i)$ in (2), the peak concentration $\mathcal{Q}_i h_p(\widetilde{d}_i)$ at the destination nanomachine corresponding to the emission by $R_i$ is obtained as

$$\mathcal{Q}_i h_p(\widetilde{d}_i) = \mathcal{Q}_i (\widetilde{d}_i)^{-3} \left(\frac{3}{2\pi e}\right)^{3/2}, \tag{4}$$

where $\widetilde{d}_i$ is distance of the destination from $R_i$. The end-to-end error rate of the cooperative molecular communication system depends on the detection performance of the individual cooperative nanomachines. The next section begins by determining the probabilities of detection and false alarm at $R_i$.

## III. Detection and Error Rate Analysis with MUI

The problem of sensing the peak concentration at the cooperative nanomachine $R_i$ in (1) can be formulated as the binary hypothesis testing problem

$$\begin{aligned} \mathcal{H}_0 &: c_i = \eta_i \\ \mathcal{H}_1 &: c_i = \mathcal{Q}_0 h_p(d_i) + \eta_i, \end{aligned} \tag{5}$$

where the null and alternative hypotheses $\mathcal{H}_0$, $\mathcal{H}_1$ correspond to the binary symbols $0, 1$ respectively. The peak concentration $c_i$ at the $i$th cooperative nanomachine corresponding to the individual hypotheses is distributed as

$$\begin{aligned} \mathcal{H}_0 &: c_i \sim \mathcal{N}(\mu_i, \sigma_i^2) \\ \mathcal{H}_1 &: c_i \sim \mathcal{N}(\mathcal{Q}_0 h_p(d_i) + \mu_i, \sigma_i^2). \end{aligned} \tag{6}$$

The optimal detection statistic $\Lambda(c_i)$ for the cooperating nanomachine $R_i$ can be obtained employing the Neyman-Pearson LRT and is given as

$$\Lambda(c_i) = \ln\left[\frac{p(c_i|\mathcal{H}_1)}{p(c_i|\mathcal{H}_0)}\right] \underset{\mathcal{H}_0}{\overset{\mathcal{H}_1}{\gtrless}} \ln\left(\frac{1-\beta}{\beta}\right). \tag{7}$$

Substituting PDFs $p(c_i|\mathcal{H}_0)$ and $p(c_i|\mathcal{H}_1)$ from (6), the logarithm of the LRT above can be evaluated as

$$\Lambda(c_i) = [-(c_i - \mathcal{Q}_0 h_p(d_i) - \mu_i)^2 + (c_i - \mu_i)^2]/2\sigma_i^2. \tag{8}$$

On further simplification, the test above reduces to

$$T(c_i) = \widehat{\alpha}_i c_i \underset{\mathcal{H}_0}{\overset{\mathcal{H}_1}{\gtrless}} \gamma, \tag{9}$$

where the parameter $\widehat{\alpha}_i$ is defined as $\widehat{\alpha}_i = \frac{\mathcal{Q}_0 h_p(d_i)}{\sigma_i^2}$ and the threshold $\gamma$ is obtained as $\gamma = \frac{(\mathcal{Q}_0 h_p(d_i))^2 + 2\mathcal{Q}_0 h_p(d_i)\mu_i}{2\sigma_i^2} + \ln\left(\frac{1-\beta}{\beta}\right)$. The result below determines the detection performance of the individual cooperative nanomachines.

*Lemma 1:* The probabilities of detection $(P_D^{(i)})$ and false alarm $(P_{FA}^{(i)})$ at the $i$th cooperative nanomachine corresponding to the test statistic $T(c_i)$ in (9) are given as

$$P_D^{(i)} = Q\left(\frac{\gamma' - \mathcal{Q}_0 h_p(d_i) - \mu_i}{\sigma_i}\right), \tag{10}$$

$$P_{FA}^{(i)} = Q\left(\frac{\gamma' - \mu_i}{\sigma_i}\right), \tag{11}$$

where $\gamma'$ is defined as $\gamma' = \gamma/\widehat{\alpha}_i$ and the function $Q(\cdot)$ is the tail probability of the standard normal random variable.

*Proof:* Given in the technical report [13, Appendix A]. ■

The cooperative nanomachines employ different types of molecules. Therefore, the joint concentration vector $\widetilde{\mathbf{c}} = [\widetilde{c}_1, \widetilde{c}_2, \cdots, \widetilde{c}_N]^T \in \mathbb{R}^{N\times 1}$ at the destination[2] corresponding to the $N$ cooperating nanomachines can be written as

$$\widetilde{\mathbf{c}} = \widetilde{\mathbf{x}} + \widetilde{\mathbf{n}}, \tag{12}$$

where the concatenated vector $\widetilde{\mathbf{x}} \in \mathbb{R}^{N\times 1}$ is defined as $\widetilde{\mathbf{x}} = [\widehat{x}_1 \mathcal{Q}_1 h_p(\widetilde{d}_1), \widehat{x}_2 \mathcal{Q}_2 h_p(\widetilde{d}_2), \cdots, \widehat{x}_N \mathcal{Q}_N h_p(\widetilde{d}_N)]^T$. Similarly, the concatenated noise vector $\widetilde{\mathbf{n}} \in \mathbb{R}^{N\times 1}$ is given as $\widetilde{\mathbf{n}} = [\widetilde{\eta}_1, \widetilde{\eta}_2, \cdots, \widetilde{\eta}_N]^T$. The sensing problem at the destination nanomachine can be formulated as the binary hypothesis testing problem

$$\begin{aligned} \mathcal{H}_0 &: \widetilde{\mathbf{c}} = \widetilde{\mathbf{n}} \\ \mathcal{H}_1 &: \widetilde{\mathbf{c}} = \widetilde{\mathbf{x}} + \widetilde{\mathbf{n}}, \end{aligned} \tag{13}$$

where $\widetilde{\mathbf{x}} = [\mathcal{Q}_1 h_p(\widetilde{d}_1), \mathcal{Q}_2 h_p(\widetilde{d}_2), \cdots, \mathcal{Q}_N h_p(\widetilde{d}_N)]^T$. Considering the logarithm of the LRT at the destination nanomachine, the decision statistic corresponding to the $N$ cooperative nanomachine transmissions is obtained as

$$\Lambda(\widetilde{\mathbf{c}}) = \ln\left[\frac{p(\widetilde{\mathbf{c}}|\mathcal{H}_1)}{p(\widetilde{\mathbf{c}}|\mathcal{H}_0)}\right] = \ln\left[\prod_{i=1}^{N} \frac{p(\widetilde{c}_i|\mathcal{H}_1)}{p(\widetilde{c}_i|\mathcal{H}_0)}\right], \tag{14}$$

where (14) follows from the fact that the peak concentrations at the destination nanomachine are conditionally independent i.e., corresponding to the alternative hypothesis $\mathcal{H}_1$ the PDF $p(\widetilde{\mathbf{c}}|\mathcal{H}_1) = \prod_{i=1}^{N} p(\widetilde{c}_i|\mathcal{H}_1)$. The individual PDFs $p(\widetilde{c}_i|\mathcal{H}_0)$ and $p(\widetilde{c}_i|\mathcal{H}_1)$ in (14) corresponding to the two hypotheses $\mathcal{H}_0$ and $\mathcal{H}_1$ are Gaussian with

$$\begin{aligned} p(\widetilde{c}_i|\mathcal{H}_0) &\sim \mathcal{N}(\widetilde{\mu}_i, \widetilde{\sigma}_i^2) \\ p(\widetilde{c}_i|\mathcal{H}_1) &\sim \mathcal{N}(\mathcal{Q}_i h_p(\widetilde{d}_i) + \widetilde{\mu}_i, \widetilde{\sigma}_i^2). \end{aligned} \tag{15}$$

Result in Lemma 2 below derives the simplified optimal Neyman-Pearson LRT at the destination nanomachine for the cooperative nano-network for moderate to high concentration scenarios with molecules ($> 10^9$).

*Lemma 2:* The optimal test statistic $T(\widetilde{\mathbf{c}})$ and the resulting test at the destination nanomachine for the binary hypothesis testing problem in (13), for a scenario in which a moderate to high number of molecules is emitted, is obtained as

$$T(\widetilde{\mathbf{c}}) = \sum_{i=1}^{N} \alpha_i \widetilde{c}_i \underset{\mathcal{H}_0}{\overset{\mathcal{H}_1}{\gtrless}} \gamma'', \tag{16}$$

where $\gamma'' = \ln\left(\frac{1-\beta}{\beta}\right) + \sum_{i=1}^{N} \theta_i$. The quantities $\alpha_i$ and $\theta_i$ are defined as

$$\alpha_i = \left[\mathcal{Q}_i h_p(\widetilde{d}_i)/\widetilde{\sigma}_i^2\right] (P_D^{(i)} - P_{FA}^{(i)}), \tag{17}$$

$$\theta_i = \left[\frac{(\mathcal{Q}_i h_p(\widetilde{d}_i))^2 + 2\mathcal{Q}_i h_p(\widetilde{d}_i)\widetilde{\mu}_i}{2\widetilde{\sigma}_i^2}\right] (P_D^{(i)} - P_{FA}^{(i)}), \tag{18}$$

with the expressions for the individual probabilities of detection $P_D^{(i)}$ and false alarm $P_{FA}^{(i)}$ for the $i$th cooperative nanomachine as obtained in (10) and (11) respectively.

*Proof:* Given in the technical report [13, Appendix B]. ■

[2]The destination nanomachine responds independently to each of the cooperative nanomachines as the various cooperating nanomachines employ different types of molecules. This is similar to the Multi-Molecule-Type Multi-Hop Network (MMT-MH) considered in works such as [17] that also employs multiple molecule types.

Employing the distributions of $\widetilde{c}_i$ for the null and the alternative hypotheses obtained in (15), together with the expression for the test statistic $T(\widetilde{\mathbf{c}})$ in (16), its distributions under the various hypotheses can be determined as

$$\begin{aligned} &\mathcal{H}_0 : \mathcal{N}\left(\sum\nolimits_{i=1}^{N} \alpha_i \widetilde{\mu}_i, \sum\nolimits_{i=1}^{N} \alpha_i^2 \widetilde{\sigma}_i^2\right) \\ &\mathcal{H}_1 : \mathcal{N}\left(\sum\nolimits_{i=1}^{N} \alpha_i\big(\mathcal{Q}_i h_p(\widetilde{d}_i) + \widetilde{\mu}_i\big), \sum\nolimits_{i=1}^{N} \alpha_i^2 \widetilde{\sigma}_i^2\right). \end{aligned} \tag{19}$$

The end-to-end probability of error for the cooperative nanonetwork follows as described in the result below.

*Theorem 1:* The end-to-end probability of error ($P_e$) at the destination nanomachine in the diffusion based cooperative molecular nano-network comprising of $N$ cooperating nanomachines, is given as

$$\begin{aligned} P_e = &\beta\left(1 - Q\left(\frac{\gamma'' - \sum_{i=1}^{N} \alpha_i(\mathcal{Q}_i h_p(\widetilde{d}_i) + \widetilde{\mu}_i)}{\sqrt{\sum_{i=1}^{N} \alpha_i^2 \widetilde{\sigma}_i^2}}\right)\right) \\ &+ (1-\beta) Q\left(\frac{\gamma'' - \sum_{i=1}^{N} \alpha_i \widetilde{\mu}_i}{\sqrt{\sum_{i=1}^{N} \alpha_i^2 \widetilde{\sigma}_i^2}}\right), \end{aligned} \tag{20}$$

where $\gamma'' = \ln\left(\frac{1-\beta}{\beta}\right) + \sum_{i=1}^{N} \theta_i$. The parameters $\alpha_i$ and $\theta_i$ are defined in (17) and (18) respectively.

*Proof:* Given in the technical report [13, Appendix C]. ■

## IV. Simulation Results

For simulation purposes, a cooperative nanonetwork is considered with diffusion coefficient $D = 10^{-6}$ cm$^2$/s. The MUI at each receiving node is modeled as a Gaussian distributed RV with mean $\mu_i = \widetilde{\mu}_i = \mu_I = 4 \times 10^{16}$ molecules/cm$^3$ that is equivalent to five interfering transmissions, each with $\mathcal{Q} = 3 \times 10^9$ molecules at a distance of 30 $\mu$m. Similar to [8], the coefficient of variation of the medium is set as 0.3 with $\sigma_i = \widetilde{\sigma}_i = 0.3\mu_I$. Fig. 2(a) shows the probability of error $P_e$ versus the number of molecules $\mathcal{Q}$ employed at each node. For simulation, the distances between the source and cooperative nanomachines, cooperative and destination nanomachines, and the source and destination nanomachines are set as $d_i = d_{sr} = 10$ $\mu$m, $\widetilde{d}_i = d_{rd} = 20$ $\mu$m, $i \in \{1, 2, 3\}$, and $d = 25$ $\mu$m respectively. It can be observed from Fig. 2(a) that the analytical values obtained using (20) coincide with those obtained from simulations, thus validating the derived analytical results. One can also observe that the end-to-end performance of the system is significantly enhanced by cooperation in comparison to the direct source-destination only communication scenario. Moreover, the end-to-end probability of error decreases progressively as the number of cooperative nanomachines $N$ increases.

Figs. 2(b) and 2(c) compare the error rate of the cooperative molecular system with that of the existing multi-input single-output (MISO) and single-input multi-output (SIMO) systems proposed in [8] for a fixed number of molecules and a fixed distance respectively. The number of cooperating nanomachines is set as $N = 2$, with two nano-transmission nodes at the information source for the MISO system, and two reception nodes at the information sink for the SIMO system. Further, the distance between each transmitting and receiving node in the SIMO and MISO systems is considered to be $d$ and 30 $\mu$m in Figs. 2(b) and 2(c) respectively. For a fair comparison, the total molecules $\mathcal{Q}$ are uniformly distributed among the transmitting nodes, i.e., $\mathcal{Q}_0 = \mathcal{Q}_1 = \mathcal{Q}_2 = \frac{\mathcal{Q}}{3}$ in the cooperative system, with $\mathcal{Q}_0 = \mathcal{Q}$ for the SIMO system, and $\mathcal{Q}_{0,1} = \mathcal{Q}_{0,2} = \frac{\mathcal{Q}}{2}$ for the MISO system. One can observe from Fig. 2(b) that the performance of the molecular communication systems with $\mathcal{Q} = 1 \times 10^9$ molecules degrades as the distance increases. Moreover, it can be seen in Figs. 2(b)-(c) that the cooperative molecular system outperforms the MISO and the SIMO systems for the scenario when the distance between the source and cooperative nanomachines is less than the distance between the cooperative and destination nanomachines. This is owing to the fact that when $d_i < \widetilde{d}_i$, the cooperative nanomachines are able to decode the source symbol correctly with high probability. Therefore, the end-to-end performance under such a scenario is dominated by the communication distance between the cooperative and the destination nanomachines. On the other hand, for the scenario with $d_i > \widetilde{d}_i$, the resulting probability of error at the cooperative nanomachines is relatively higher, leading to erroneous transmissions to the destination nanomachine. Hence, the performance of the cooperative system is dominated by the source-cooperative nanomachine distances. In contrast to the SIMO system with $\mathcal{Q}_0 = \mathcal{Q}$ molecules, the cooperative scenario uses $\mathcal{Q}_0 = \frac{\mathcal{Q}}{3}$ molecules at the source. Therefore, the cooperative system has a higher probability of error in comparison to the SIMO system. However, the cooperative system outperforms the MISO system for this scenario as well. Moreover, a significant improvement in the error performance is attained for the scenario when the cooperative nanomachines are at equal distances from both the source and destination nanomachines. Due to space limitations, additional simulation results for detection performance at the destination nanomachine are presented in the technical report [13].

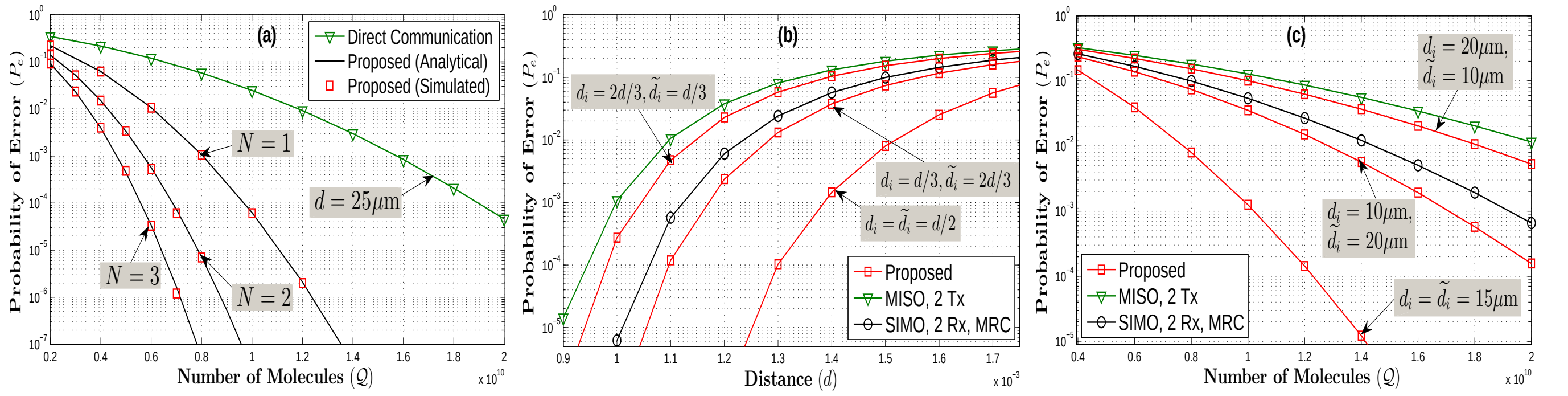

Fig. 2. Error rate of the diffusion based cooperative molecular communication system for various scenarios with (a) direct communication (b) MISO and SIMO systems with fixed number of molecules (c) MISO and SIMO systems with fixed distance.

## V. Conclusion

This work presented a performance analysis for diffusion based cooperative molecular communication systems wherein multiple nanomachines cooperate with the source to enhance the end-to-end reliability in the presence of MUI. Analytical expressions were derived for the optimal test statistics at the cooperative, destination nanomachines, together with the resulting probabilities of detection, false alarm as well as the end-to-end error rate. Moreover, the proposed system was seen to yield a performance improvement in comparison to the existing SIMO and MISO systems.